# Realization of super-bunching laser with giant high-order correlations and extreme multi-photon events

Chengbing Qin[1,2,3,*,†], Yuanyuan Li[1,2,†], Yu Yan[3,†], Jiamin Li[3,*], Xiangdong Li[1,2], Yunrui Song[1,2], Xuedong Zhang[1,2], Shuangping Han[3], Zihua Liu[3], Yanqiang Guo[3,4], Guofeng Zhang[1,2], Ruiyun Chen[1,2], Jianyong Hu[1,2], Zhichun Yang[1,2], Xinhui Liu[1,2], Liantuan Xiao[1,2,3,4,*], and Suotang Jia[1,2]

1. State Key Laboratory of Quantum Optics and Quantum Optics Devices, Institute of Laser Spectroscopy, Shanxi University, Taiyuan, Shanxi 030006, China
2. Collaborative Innovation Center of Extreme Optics, Shanxi University, Taiyuan, Shanxi 030006, China
3. Department of Physics and Optoelectronics, Taiyuan University of Technology, Taiyuan, 030024, China
4. Key Laboratory of Advanced Transducers and Intelligent Control System of Ministry of Education, Taiyuan University of Technology, Taiyuan, 030024, Shanxi, China

[*] Corresponding authors. Email:

Chengbing Qin, chbqin@sxu.edu.cn;

Jiamin Li, lijiamin@tyut.edu.cn;

Liantuan Xiao, xlt@sxu.edu.cn.

† These authors contributed equally to this work.

**Abstract**

Non-classical light sources emitting bundles of *N*-photons have essential applications ranging from fundamental tests to quantum information processing. However, high-order correlations, $g^{(N)}(0)$ ($N \geq 2$), are still limited to hundreds. Herein, we report the realization of a super-bunching laser via stochastic nonlinear interactions in an optimized photonic crystal fiber, with $g^{(2)}(0)$ and $g^{(5)}(0)$ reaching $5.86\times10^4$ and $2.72\times10^8$. This laser presents extreme photon-number fluctuations and ubiquitous multi-photon events, with the occurrence of 31 photons from a single pulse at mean photons of $1.99\times10^{-4}$. The probability of this extreme event has been enhanced $10^{139}$ times compared to a coherent laser with a Poissonian distribution. To show the advantage of extreme multi-photon events, we performed *N*th-order correlated imaging by the coincidence of *N* photons in a single pulse, which can resist stray light up to $10^4$ times stronger than the super-bunching photons. Remarkably, the noise can be wholly eliminated with $N \geqslant 6$, and the rejection of stray noise in our current technology is only limited by the maximum count rates of the single-photon detectors. Our study showcases the ability to achieve non-classical laser with giant high-order correlations and extreme multi-photon events, paving the way for high-order correlation imaging, extreme nonlinear optical effects, hyperspectral image, and multi-photon physics.

## 1. Introduction

Quantum optics was born from the study of photon statistics. The statistical properties of a light beam can be characterized using the second-order correlation function, $g^{(2)}(\tau)$, first proposed by Hanbury Brown and Twiss[1,2]. Shortly after, Glauber introduced a quantum mechanical description of the second-order correlation function[3]. The normalized $g^{(2)}(0)$, an essential tool for quantum state classification, aiming at recognizing what kind of photon number distributions the investigated light source emits, can be distinguished into sub-Poissonian [$g^{(2)}(0)<1$], Poissonian [$g^{(2)}(0)=1$], and super-Poissonian [$1<g^{(2)}(0)$] distributions[4-6], which are also known as anti-bunching, coherent, and bunching emissions, respectively. Super-bunching light sources with $g^{(2)}(0)>2$ indicate the existence of correlated photons or a multi-photon bundle emission and also suggest extreme photon-number fluctuations stronger than those of thermal light. Such highly non-classical states of light are of fundamental interest to quantum optics and constitute a starting point for engineering more complex quantum states. They are also essential for newly emerging, more resource-efficient photonic architectures for universal quantum computation and quantum error correction using individual, higher-dimensional systems, as well as for the optimal capacity of a quantum communication channel and Heisenberg limited quantum metrology (interferometry)[7-13]. Furthermore, the efficiency of multi-photon nonlinear light-matter interactions, such as two-photon absorption, is proportional to the bunching peak $g^{(N)}(0)$ of the light field, which has been experimentally verified with both thermal light and entangled photon pairs[14-16]. Therefore, the super-bunching effect of light fields is of essential importance not only from the viewpoint of fundamental physics but also for applications such as quantum communications, computation, and nonlinear optics. Although the super-bunching effect has been observed in various systems, such as parametric down-conversion[17,18], quantum dots[19,20], atomic ensembles[21], generic tunnel junctions[22], and superconducting circuits[23], the existing super-bunched photon sources generally suffer from either low quantum yields, small $N$-photon bundle emission, weak photon number fluctuations, or low super-bunching index values ($g^{(2)}(0)$ limited to hundreds)[24].

In this study, we realized a giant super-bunching laser, with $g^{(2)}(0)$ reaching $5.86\times10^4$ in the visible region, through the nonlinear interactions between a low-power pump laser and an optimized photonic crystal fiber (PCF). High-order correlation functions of the laser were also determined, with $g^{(5)}(0)$ reaching $2.72\times10^8$, at least six orders of magnitude higher than that of the thermal light[25,26]. The measurement of the photon number probability distribution, uncovered by a home-made photon-number-resolving single-photon counting module (SPCM) array, demonstrated that the photon statistics of the super-bunching laser with a giant $g^{(2)}(0)$ were far beyond the coherent (Poissonian) and thermal (Bose-Einstein) distributions and presented extreme photon-number fluctuations. The $N$-photon production probability, scaling as the $N$th-order normalized correlation functions, was dramatically enhanced. Compared with coherent light, the 31-photon production probability could be improved $10^{139}$ times with a mean number of photons per pulse ($\langle n\rangle$) of $1.99\times10^{-4}$. Counterintuitively, with the decrease in $\langle n\rangle$, the likelihood of bundled photon emission with an enormous $N$ value will increase rather than decline. To show the advantage of this non-classical super-bunching laser, we performed 2$^{nd}$- to 20$^{th}$-order correlated imaging; almost all the noise can be eliminated for the correlated imaging even when the photon ratio between the stray noise and the super-bunching laser exceeded $10^4$. The results confirmed that our high-order correlated imaging protocol can reject all uncorrelated events that arise from either sensor noise or unwanted background illumination.

## 2. Results

### 2.1 *Experimental setup.*

The core concept for realizing the super-bunching effect is to enable light emission via stochastic processes rather than coherent resonance. Thus, we implemented the super-bunching laser through a bright squeezed vacuum light with an optimized PCF through complex and stochastic nonlinear interactions, as the experimental layout depicted in Fig. 1a (see Methods for details). A semiconductor saturable absorber mirror (SESAM) mode-locked fiber laser acted as the seed pulse, successively amplified by a pre-amplifier and a main amplifier. The super-bunching laser was generated by injecting a low-power picosecond pump laser into the optimized PCF, which was composed of five layers of periodically arranged air holes, forming a ring-shaped configuration within the fiber (see Fig. S1 for the morphology and dispersion curves). The high-order correlation functions and photon number probability distribution were measured by an array of silicon-based SPCMs and a multi-channel event timer (MCET) with a time-correlated single-photon counting (TCSPC) unit.

### 2.2 *Measurement of $g^{(N)}(0)$*.

The spectral region and the super-bunching effect could be varied by changing the pump power of the main amplifier. Fig. 1b presents the input-output power curve for this laser with a center wavelength of 605 nm, a bandwidth of 70 nm, and a repetition of 80 MHz. The threshold ($P_{th}$) for the 605 nm emission was determined to be 1.65 W. As expected and reported previously[27], the value of $g^{(2)}(0)$ for the emission beyond the threshold (*e.g.*, the pump power was three folds of the threshold, $P_{pump}=3P_{th}$) was close to 1, the characteristic feature of coherent lights. As the pump power approached the threshold, $g^{(2)}(0)$ tended to 2, the characteristic feature of chaotic light. Surprisingly, the value of $g^{(2)}(0)$ surpassed 2 and favored thousands with the decrease in the pump power (Fig. 1c), and the observed maximum $g^{(2)}(0)$ reached up to $5.86\times10^4$ (Fig. S6). Fig. 1d presents a typical second-order correlation function with $g^{(2)}(0)$=16747, which is two orders larger than the reported values[28-30], indicating their remarkable non-classical effects. These results declare that the light field can be substantially tailored from a coherent to a super-bunching light field. Furthermore, higher-order correlations may provide a detailed understanding of the nature of quantum many-particle states. For example, $g^{(3)}$ and $g^{(4)}$ can provide information about the skewness and kurtosis of the fluctuation statistics, respectively[31]. As shown in Fig. 1e, $g^{(3)}(0)$, $g^{(4)}(0)$, and $g^{(5)}(0)$ reached $9.54\times10^4$, $1.29\times10^6$, and $2.72\times10^8$, respectively, under an intermediate condition with $g^{(2)}(0)$=75 and $\langle n \rangle$ at the $10^{-2}$ level. The rising tendency for the super-bunching laser followed $e^{5N}$, indicating that $g^{(31)}(0)$ could reach $10^{67}$. Note that the measurement of $g^{(N)}(0)$ with $N$>5 was restricted by the exponential growth of the photon numbers and the limited computer power, rather than the experimental scheme itself. We also demonstrated the high-order correlation functions of the coherent and thermal light. As expected, the value of $g^{(N)}(0)$ remained at unity for the coherent light source, and that for the thermal light was close to $N$!, agreeing well with the result of the Bose–Einstein distribution (Fig. S7). The value of $g^{(5)}(0)$ for the super-bunching light source was enhanced by at least six orders of magnitude compared with that of the thermal light, which further manifested the pronounced non-classical effects of our super-bunching laser. Another characteristic feature of the super-bunching laser was that the value of $g^{(N)}(0)$ increased with the decrease in $\langle n \rangle$ (at the same pumping power while using an attenuator, *i.e.,* neutral density filters, to vary $\langle n \rangle$). Note that the value of $g^{(2)}(0)$ improved from 49 to 2284 when $\langle n \rangle$ decreased from $9.44\times10^{-3}$ to $8.68\times10^{-5}$ (Fig. 1f). These results were completely different from those of the coherent and thermal light with the same $g^{(N)}(0)$ value under different $\langle n \rangle$ values (see Figs. S5-S11).

### 2.3 ***Photon statistics.***

To uncover the underlying mechanism of the giant $g^{(N)}(0)$, we obtained the photon number probability distribution of the super-bunching laser with 31 SPCMs under different conditions. To check the experimental system, we first performed photon statistics on the conventional pulsed laser. The experimental results were consistent with the theoretical predictions (Fig. S5). For the super-bunching laser, when the pump power was far above the threshold (*e.g.*, $P_{\text{pump}}=3P_{\text{th}}$), the value of $g^{(2)}(0)$ exactly equaled 1, and their photon statistics with $\langle n \rangle$ ranging from $1.37\times10^{-1}$ to $1.97\times10^{-4}$ agreed well with the Poissonian distribution, as shown in Fig. 2g. However, when the pump power was close to and especially below the threshold, $g^{(2)}(0)$ had a giant value, and the photon number probability distributions dramatically deviated from the Poissonian distribution. Even with an $\langle n \rangle$ of $4.70\times10^{-4}$, the probability of a 31-photon bundled emission was still up to $10^{-11}$ (Fig. 2a). To quantify this deviation, we define the ratio $\zeta_{ij}(N)$ as

$$\zeta_{ij}(N)=\frac{P_i(N)}{P_j(N)} \tag{1}$$

where $P_i(N)$ is the probability of an $N$-photon bundled emission from the $i$-type light source within a single pulse. For example, $P_C(31)$ can be understood as the probability of a 31-photon bundled emission within a single pulse from a coherent laser, and $\zeta_{SC}(N)$ is the ratio of $P(N)$ between a super-bunching and a coherent laser. It can be demonstrated that, with $g^{(2)}(0)=1$, the values of $\zeta_{SC}(N)$ were limited to values in the range of 0.3–3 (Fig. 2h), which was mainly due to the limited acquisition time and thus the measurement errors. While with $g^{(2)}(0)$ improving to 347 ($\langle n\rangle=1.89\times10^{-1}$), the value of $\zeta_{SC}(31)$ reached up to $10^{53}$ and presented extreme photon-number fluctuations, as shown in Fig. 2d. The maximum $\zeta_{SC}(N)$ value in the experiment reached up to $10^{139}$ with an $\langle n\rangle$ of $1.99\times10^{-4}$ and $g^2(0)$ of 3859, as shown in Fig. S6.

More intriguingly, even at the same pump power, $g^{(2)}(0)$ and $\zeta_{SC}(N)$ presented dramatical enhancement synchronously with the decrease in $\langle n\rangle$, as shown in Fig. 2d. For example, $\zeta_{SC}(31)$ was improved from $2.13\times10^{73}$ to $1.50\times10^{126}$ as $\langle n\rangle$ decreased from $4.01\times10^{-2}$ to $4.70\times10^{-4}$, and $g^{(2)}(0)$ was enhanced from 1562 to 11637 (Fig. S9). These results indicated that small-probability events might always occur in this case and thus manifest extreme photon-number fluctuations. To illustrate this conclusion, we plotted the measured sequences of the multi-photon events of the super-bunching laser with $g^{(2)}(0)=4538$ and $\langle n\rangle=1.11\times10^{-3}$ in 1 s, as depicted in Figs. S13–S16. We found that the multi-photon events could be easily distinguished, and the extreme events continued to occur, even with the photons exceeding their mean values ($N/\langle n\rangle$) of more than $2.8\times10^4$. For comparison, we also plotted the photon sequences of the coherent laser with $\langle n\rangle=4.94\times10^{-3}$, where only three 3-photon bundled emissions occurred in 1 s. Fig. 2b presents the variations of $\langle n\rangle$ and $P(N)$ with the pump power. As expected, these two values intuitively increased with the rise of the pump power (the maximum pump powers were slightly larger than $P_{\mathrm{th}}$). In contrast, the values of $\zeta_{SC}(N)$ presented the reverse variation trend, *i.e.*, $\zeta_{SC}(N)$ decreased with the rise of the pump power. This phenomenon hinted that the weaker the pump power was, the higher the $N$-photon bundle emission generation probability became. Note that when the pump power was much lower than $P_{\text{th}}$, both $P(N)$ and $g^{(2)}(0)$ were close to the coherent light due to the extremely weak photon emissions, which were close to Gaussian noise (Figs. S17). Another feature of the super-bunching light source was that $P(N)$ and $\zeta_{SC}(N)$ presented the same tendency for similar $\langle n \rangle$ values but rather different pump powers. With $\langle n \rangle$ close to $1.00\times10^{-2}$ as an example (Figs. 2c and 2f), the $\zeta_{SC}(N)$ curves at four pump powers overlapped. Furthermore, we found that the probability of a multi-photon bundled emission (mainly with $N$ between 10 to 25) showed a plateau rather than the "heavy-tail" probability distributions reported previously[30,32], as presented in Figs. 2b and 2c. Counterintuitively, $P(N>25)$ was more extensive than $P(N=25)$ in most cases, suggesting the divergent feature of the probability distributions. To explore this phenomenon, we selected a convergent probability distribution and performed photon statistics with arbitrary $M$ channels ($M\leq31$). The results are shown in Fig. S18. "Upturned-tail" distributions were observed when the number of SPCM channels decreased. Hence, the "upturned-tail" distributions principally originated from $N$-photon bundle emissions with $N>M$. This result advised that the multi-photon bundle emission exceeded 31 in our experiment.

To further illustrate the extremely fluctuating feature of the super-bunching laser, we also obtained the photon number distribution of thermal light, as shown in Fig. 2i, which agreed closely with Bose-Einstein statistics (Fig. S12). Moreover, the $g^{(2)}(0)$ value for the thermal light was close to 2. These two features agreed well with the theoretical predictions. The ratios of the $P(N)$ values between the thermal and coherent light, $\zeta_{TC}(N)$, increased exponentially with the increase in the photon number for small $\langle n \rangle$ values, due to the "heavy-tailed" distributions in thermal light sources. Although $\zeta_{TC}(N)$ exhibited a similar tendency, the maximum value was limited to $10^{12}$, at least 30 orders of magnitude smaller than that of the super-bunching laser with a similar $\langle n \rangle$ value.

## 2.4 ***High-order correlated imaging.***

To show the advantage of this super-bunching laser, we performed high-order correlated imaging in the presence of intense stray light noise. The experimental configuration is shown in Fig. 3a. The laser beam illuminated the target object (three letters "SXU"). Then, the scattering light was collimated and collected by a fiber bundle, in which each fiber guided photons to an SPCM. A thermal light was used to simulate real-world conditions and intense stray noise. The thermal light overlayed the super-bunching beam through a beam splitter (BS) and was detected by the same SPCM array. The $N$th-order correlated imaging was demonstrated by calculating the $N$th coincidence events from arbitrary $N$ SPADs. The imaging was determined by raster scanning the Galvo mirror, as depicted in Fig. 3b. Here, $\langle n \rangle$ for the super-bunching laser was $3.54\times10^{-4}$ while that for the noise was 4.4, *i.e.*, the ratio of photon numbers between the noise and the super-bunching laser exceeding $10^4$. It can be found that for the single photon and $2^{nd}$-correlated imaging (*i.e.*, 2-photon correlated imaging shown in Fig. $3b_1$ and $3b_2$), the signals (object) are fully submerged in the background noise. For $3^{rd}$- to $5^{th}$-correlated imaging, the object can be well distinguished, and the signal-to-noise ratio (SNR) has dramatically improved. For even higher-order correlated imaging ($N\geq6$), the noise has been completely removed (Figs. $3b_6$ and S19).

These results can be well understood from the photon number distributions with and without super-bunching laser (*i.e.*, pure stray noise), as presented in Figs. 3c. For stray noise, the 5-photon events are only $7.6\times10^{-3}$ Hz when calculating $5^{th}$ coincidences from arbitrary 5 SPCMs. Even calculating all the 20 SPADs, the 6- and 7-photon events are only 1.52 Hz and 0.029 Hz for thermal light (Figs. S20). In contrast, the 5-photon events are 18.4 Hz for the signal with both super-bunching laser and noise. Furthermore, 6- to 20-photon events also show remarkable values for the super-bunching laser. We also calculated the *SNR* for the $N$th-order correlated imaging, as illustrated in Fig. 3d.

Note that the variation of *SNR* as a function of correlated order is in good agreement with the correlated imaging. With *N* ranging from 3 to 20, the mean number of photons per pixel, $\tilde{n}$, for the correlated imaging varied gradually from 12.13 to 5.11 Hz (with an integral time of 0.5 s), as presented in Fig. 3e. These results indicated that the rejection of stray noise in our current technology can be further improved, which is only limited by the maximum count rates of the single-photon detectors. To eliminate the possible influence of the combinations from arbitrary SPAD channels, we also presented 20 combinations of $19^{\text{th}}$-correlated imaging, as shown in Fig. S21, and no significant difference between them can be observed. The value of $\tilde{n}$ remains for the combination of arbitrary channels, as illustrated in Fig. 3f. This phenomenon indicates that high-quality imaging can be demonstrated by using sufficient combinations. Particularly, for high-order correlated imaging with $N \geq 5$, there are $10^6$ combinations ($\sum_5^{20} C_{20}^i = 1042380$), meaning that by using all these combinations, the integral time can be reduced from 0.1 s to 0.1 μs with similar $\tilde{n}$. In this case, high-quality imaging can be demonstrated with extremely weak photon flux but exceptionally short integral time (100 ns per pixel) under huge noise.

## 3. Conclusion

In this study, we provided an experimental demonstration of a non-classical laser with giant high-order correlations and an extremely large probability for multi-photon events through the stochastic nonlinear interactions between a picosecond laser and a PCF. Both the second-order correlations and photon statistics of this light source can be tailored from Poissonian distributions to super-bunching distributions by varying the power of the pumping laser and/or the mean photons per pulse using neutral density filters. The unparalleled characteristics of the super-bunching laser are especially important for multi-photon correlated imaging, the nonlinear optics of fragile structures, and quantum thermodynamics. For example, the super-bunching laser can be used for extremely weak light detection and ranging as well as communications. Although extremely small $\langle n \rangle$ values will be detected in this case, the accuracy of the information can be determined, and the intense background noise can be eliminated based on multi-photon-correlated calculations or $N$-photon heralding measurements. The extremely high probabilities of multi-photon events are already suitable for practical applications. The occurrence of multi-photon events also results in extremely fluctuating light and offers much more efficiency for multi-photon effects than coherent sources with the same mean number, which has been proven by experiments[15,32]. The super-bunching laser with giant $g^{(N)}(0)$ values may enhance the multi-photon effects by several orders of magnitudes. Furthermore, the super-bunching laser manifests giant high-order correlations for the full broad wavelengths from visible to near-infrared wavelengths (Fig. S7). This feature is also beneficial for hyperspectral imaging without any spectrometers[33,34], only through cross-correlation with different wavelengths.

## Figure Captions

Figure-1

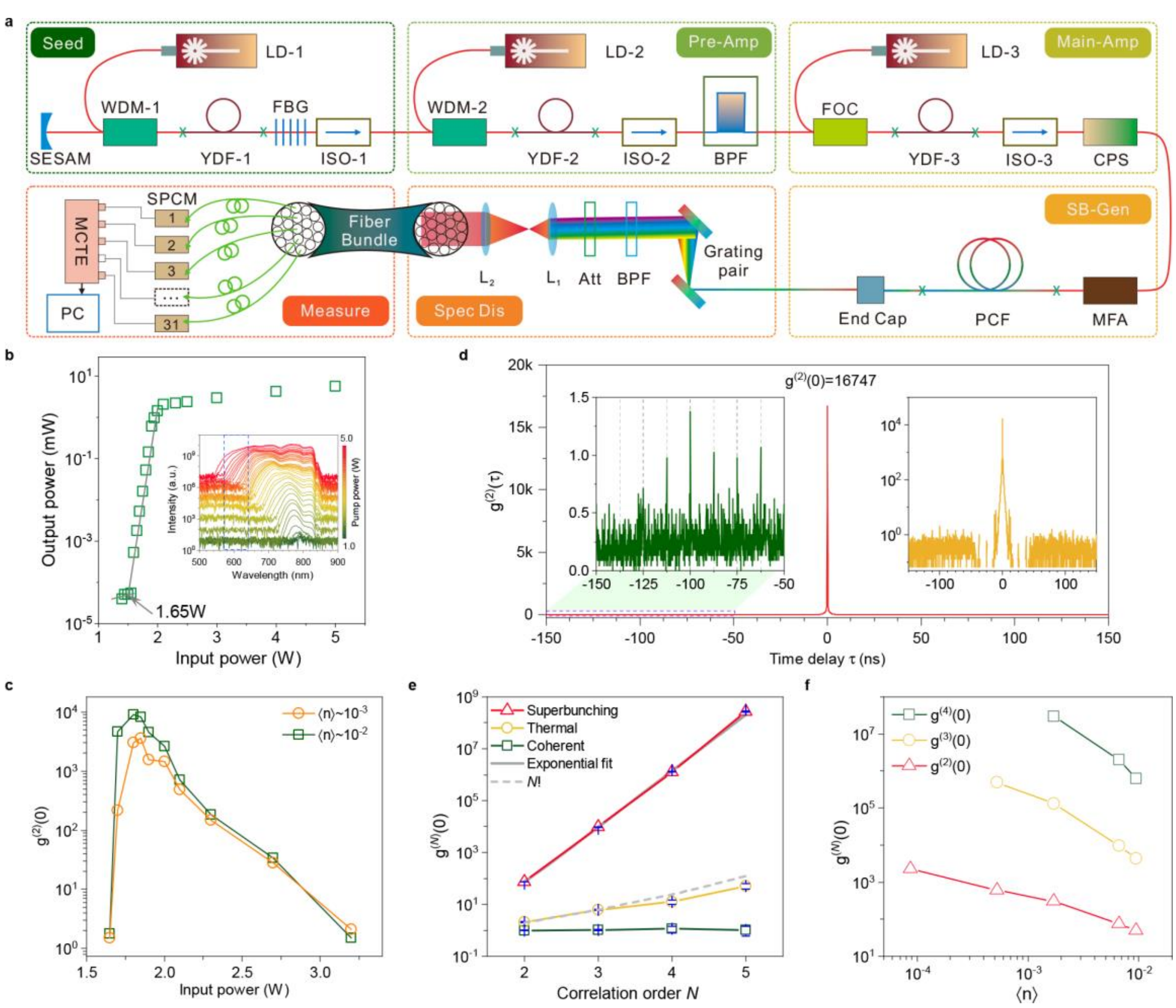

**Fig.1 Realization and measurement of high-order correlation of the super-bunching laser. a,** Schematic of the realization and measurement of the super-bunching laser. The features of the super-bunching laser could be varied by tuning the pump power of the main amplifier. See Methods for details. **b,** Output power of the super-bunching laser versus the input pump power of the main amplifier. The inset shows the spectrum evolution in the visible region as a function of the pump power. **c,** Measured second-order correlation function, $g^{(2)}(0)$, as a function of the pump power for two mean numbers of photons per pulse, $\langle n \rangle$. **d,** Second-order correlation function as a function of the time delay ($\tau$), $g^{(2)}(\tau)$, with $g^{(2)}(0)$=16747. The insets show $g^{(2)}(\tau)$ with a logarithmic scale and an enlargement of $g^{(2)}(\tau)$, with $\tau$ between −150 and −50 ns. From the enlargement, bunched and normalized peaks at $\tau=NT$ can be observed with $T=1/f=$ 12.5 ns and $N$ as integer values. **e,** Measured high-order correlation function $g^{(N)}(0)$ of the super-bunching laser compared with the coherent laser and thermal light. The gray solid line is an exponential fit of $g^N(0)$ with the form $11.48\times e^{5N}$. The short-dashed line is the result of $N!$. **f,** $g^N(0)$ of the super-bunching laser as a function of $\langle n\rangle$. The pumping powers for **e** and **f** were 1.85 W. The center wavelength and bandwidth of the super-bunching laser for these measurements were 635 and 70 nm (as the dashed rectangle shown in **c**) with a repetition frequency of 80 MHz, respectively.

Figure-2

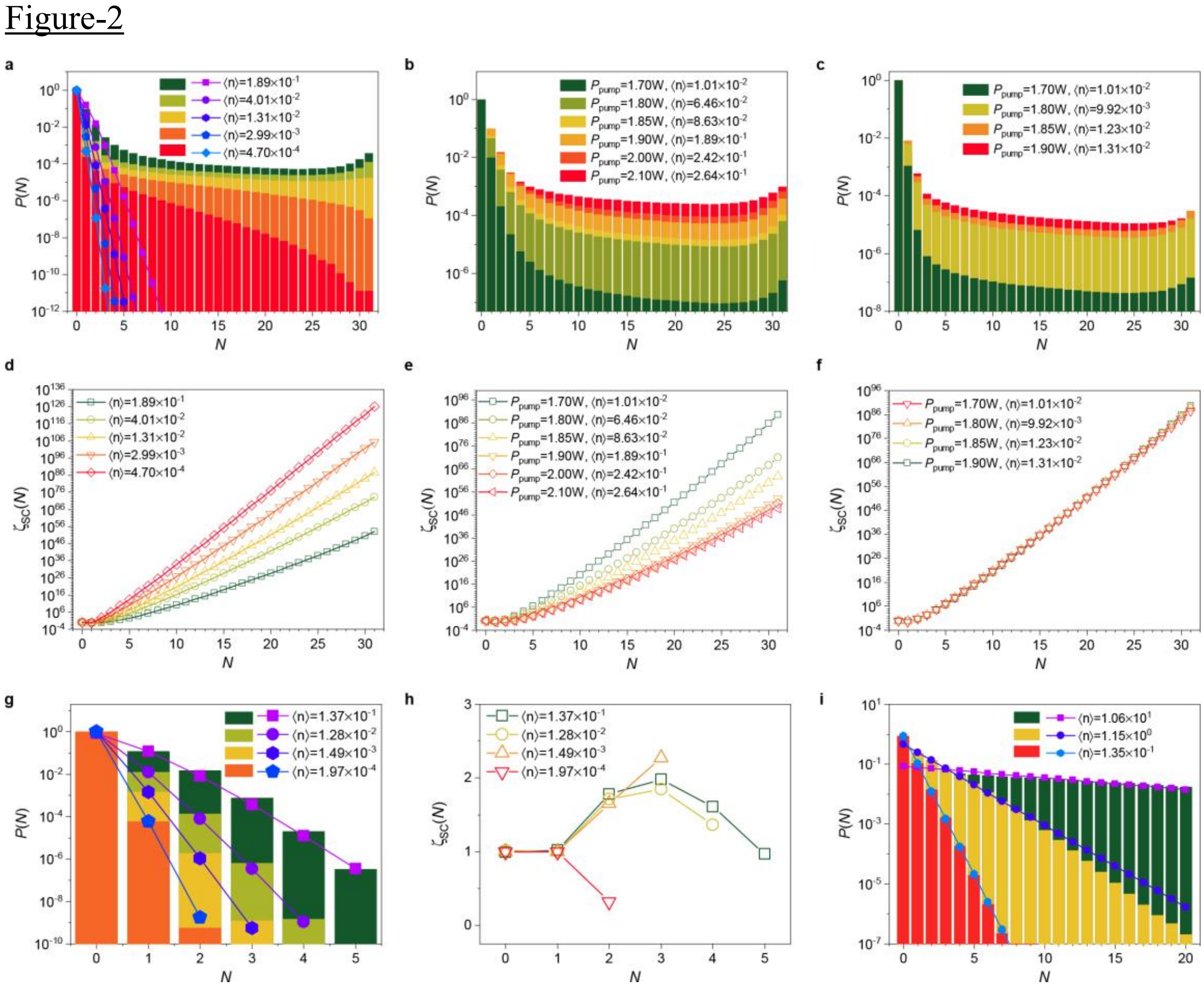

**Fig. 2 Measurement of the photon number probability distribution, *P*(*N*). a,** $P(N)$ of the super-bunching laser as functions of $\langle n\rangle$, at a pump power of 1.85 W. The mean number $\langle n\rangle$ was varied using neutral density filters. The solid symbols represent the Poissonian distributions with the same $\langle n\rangle$ value. With the decrease in $\langle n\rangle$, the $g^{(2)}(0)$ values are 347, 1562, 2811, 4573, and 11637 (see Fig. S9). **b,** $P(N)$ at different pump powers. The $\langle n\rangle$ value under each pump power is denoted. With the increase of pump

power, the $g^{(2)}(0)$ values are 215, 288, 595, 347, 191, and 145 (see Fig. S10). **c,** $P(N)$ with similar $\langle n \rangle$ values but different pump powers. With the increase of pump power, the $g^{(2)}(0)$ values are 215, 3024, 3609, and 2811 (see Fig. S11). The corresponding ratios of $P(N)$ between the super-bunching and coherent lasers ($\zeta_{SC}(N)$) in **a**, **b**, and **c** are shown in **d**, **e**, and **f**, respectively. **g,** $P(N)$ of the super-bunching laser with various $\langle n \rangle$ values at $P_{\text{pump}}$= 4.5 W and $g^{(2)}(0)$=1.05 (see Fig. S5). The solid symbols represent the Poissonian distributions. **h,** Corresponding $\zeta_{SC}(N)$ for the super-bunching laser in **g**. **i,** $P(N)$ of the thermal light with various $\langle n \rangle$ values and $g^{(2)}(0)$~2.0. The solid symbols are the simulated Bose-Einstein distributions.

Figure-3

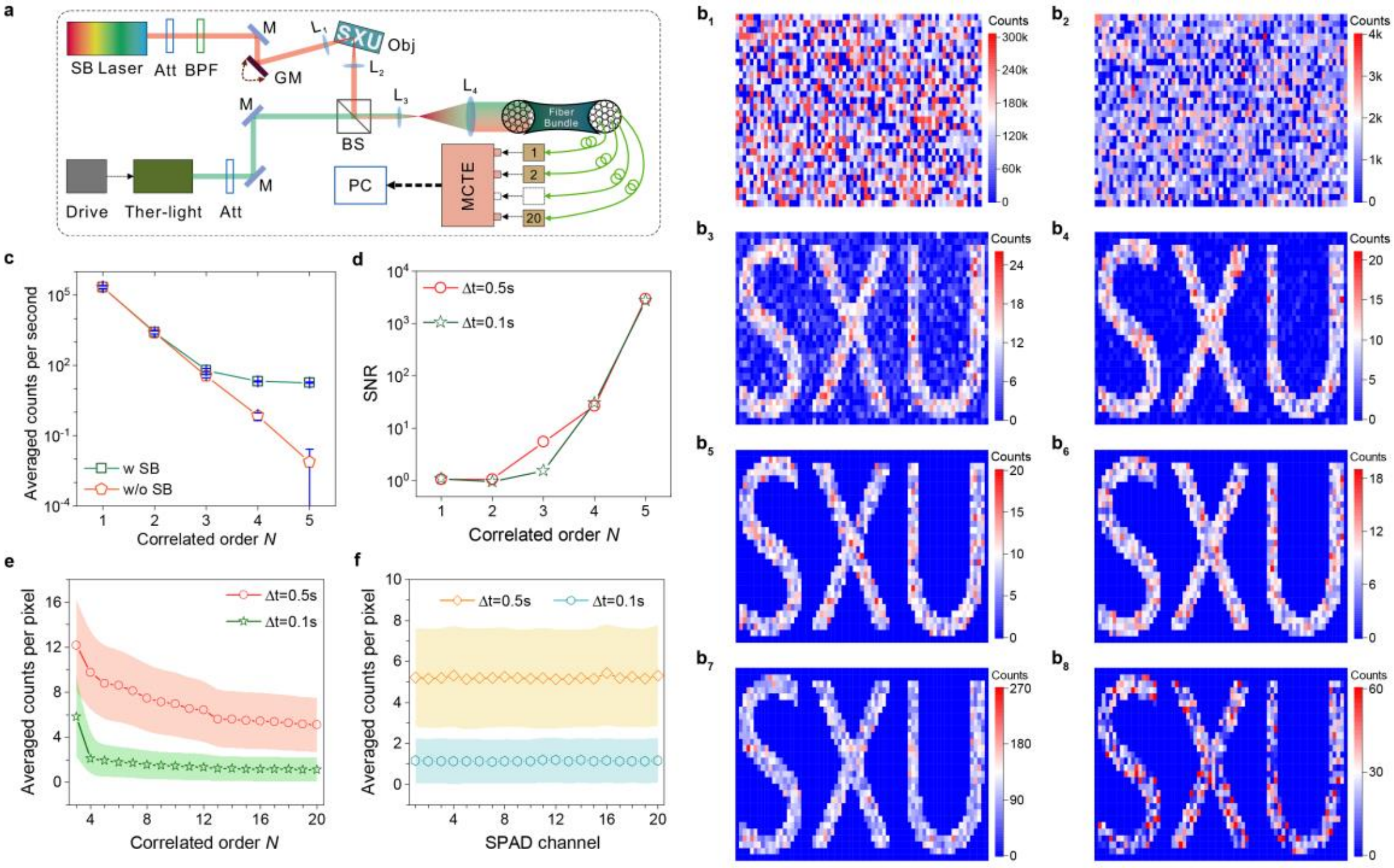

**Fig. 3 High-order correlated imaging in the presence of intense stray light. a**, Schematic of the experimental setup for the high-order correlated imaging. See Methods for details. The super-bunching laser illustrates the object (three hollow letters "SXU"), and the background noise originates from a thermal light source. The photon ratio between the noise and the laser reaches 12420. **b**, $N$th-order correlated imaging with $N$ ranging from 1-5 (**b$_1$** to **b$_5$**), and 20 (**b$_6$**), respectively. The integral time for each pixel was 0.5s. **b**$_7$ and **b**$_8$ are the summary of 20 combinations for 19$^{th}$-correlated imaging, with the integral time for each pixel being 0.5s and 0.1s, respectively. For 5$^{th}$-correlated imaging, only 3 pixels occur noises. For larger-order correlated imaging, the noise is completely removed. **c**, Averaged *N*-photon events per second calculated from arbitrary N SPCMs with (w) and without (w/o) the super-bunching laser. The total acquisition time was 1000 s. The repetition frequency was 5 MHz. Without the super-bunching laser, the maximum *N* is only 7 when calculating all 20 SPCMs, while with the super-bunching laser, the maximum *N* is up to 20 (Fig. SX). **d**, The single-to-noise ratio (SNR) of *N*th-order correlated imaging under two integral times. **e**, The averaged number of photons per pixel ($\tilde{n}$) as a function of correlated order (*N*) under the integral time of 0.5s and 0.1 s, respectively. $\tilde{n}$ for single- and two-photon imaging have been wiped out. **f**, $\tilde{n}$ for 19$^{th}$-correlated imaging under different combinations under two integral times. The shadow areas in **e** and **f** are the measurement errors.

## Methods

### 1. Generation of the super-bunching light sources

The experimental implementation for generating the super-bunching light source is illustrated in Fig. 1a. The seed laser was a semiconductor saturable absorber mirror (SESAM) based mode-locked fiber laser, generated by a semiconductor laser diode (LD-1), a SESAM, a wavelength division multiplexing (WDM-1), a fiber Bragg grating (FBG), a fiber isolator (ISO-1), and a piece of 0.5 m long 4/125 μm single mode Ytterbium-doped optical fiber (YDF-1), respectively. The semiconductor laser has a central wavelength of 980 nm and a maximum pumping power of 600 mW. The modulation depth, saturation fluence, and relaxation time constant of the SESAM are 14%, 60 $\mu J/cm^2$, and 3 ps, respectively. The FBG exhibits reflectivity of 10% with a spectral bandwidth of 0.3 nm at the center wavelength of 1064 nm and acts as the output cavity mirror. The seed laser was locked when the pump power of LD-1 reached 75 mW, resulting in a pulsed laser with a center wavelength of 1064 nm, pulse width of about 100 ps, repetition rate of 80 MHz, and power of 21 mW. After passing through the ISO-1, the laser pulse was successively amplified by one pre-amplifier and one main amplifier. The pre-amplifier, employing a piece of 0.5 m YDF (YDF-2), was pumped by 600 mW fiber pigtailed 980 nm laser diode (LD-2). When the pump power of LD2 reached 220 mW, the pre-amplified pulse was 82 mW. The pre-amplified pulse is unidirectional transmission by another isolator (ISO-2), and the amplified spontaneous emission (ASE) is filtered by a band-pass fiber filter (BPF). The main amplifier was constructed by a piece of 3 m YDF (YDF-3) with a core/inner diameter of 10/125 μm, and pumped by a 10 W fiber pigtailed 980 nm (LD-3) *via* a 2×1 fiber optical combiner (FOC). One-way transmission is maintained through a fiber isolator (ISO-3). A cladding power stripper (CPS) was used to remove the residual laser from the optical fiber cladding. When the pumping power of LD-3 reached 3 W, the pulse laser with the maximum power of 500 mW and a wavelength center of 1064 nm was used to pump the optimized photonic crystal fiber (PCF) to generate the super-bunching light source.

A mode field adapter (MFA) was used to match the mode field area between the output pigtail of the main amplifier and the PCF. The input pigtail is a double-clad passive fiber, and the output pigtail is a single-mode fiber. The structure of PCF is composed of five layers of circular air holes. The diameter of all air holes is 1.9 μm, and the distance between two holes is 3.3 μm, as illustrated in the inset of Fig. S1. In this experiment, the pumping powers of LD1 and LD2 were fixed as 75 mW and 220 mW. The output power and the unique features of the super-bunching light were varied by tuning the pumping power of LD3. The generated super-bunching light was collimated to a pair of grating to shape the light beam. The interested wavelength regions can be selected by BPFs. In this experiment, we mainly focused on the super-bunching light source with a central wavelength of 605 nm and width of 70 nm by using a long pass filter (LPF, Chroma, ET570LP) and a short pass filter (SPF, Semrock, FF01-650/SP-25), the transmission spectra of the two filters are presented in Fig. S1. The intensity or the mean photons per pulse ($\langle n \rangle$) were attenuated (Att) by neutral density filters (JCOPTIX, OFR1-M1). After attenuation, the light beam was expanded by two lenses ($L_1$ and $L_2$) and collimated to a microlens array (8×8) with a distance between each element of 0.5 mm. Each microlens is coupled with a multimode fiber with a core/cladding diameter of 550/125 μm to form a fiber bundle (Fig. S1). By adjusting and aligning the beam expander, the super-bunching light beam was coupled to fiber arrays. Then, we selected 31 fibers with the front intensities to couple a single photon counting modules (SPCM) array with 31 detectors. These signals, including the arriving time of each photon, were then converted to a personal computer through a high-through multi-channel event timer (MCTE, 32 channels) with a time-correlated single photon counting unit (TCSPC) (Seru, TCU200-32). The time resolution and the total sustained count rate of this MCTE are about 20 ps and 90 Mcps (million counts per second). Considering the heterogeneity of the light beam as well as the inhomogeneous coupling and detection efficiency of each fiber and SPCM, the photon counting per second for each SPCM shows a quiet difference, especially for extremely small $\langle n \rangle$, as presented in Fig. 1e for example. The photon statistics and the corresponding high-order correlation function were calculated by a homemade program.

## 2. Generation of the pseudo-thermal light

To prove the reliability of the experimental system and check the photon statistics of thermal lights, we perform the photon number probability distribution of a pseudo-thermal light, which was generated by a micro-disk laser (LDRVMINI) derived by random noise and modulated by an acousto-optic modulator (AOM, MT110-A1, 5-VIS) with varied frequencies and duty cycles, as illustrated in Fig. S2. The linewidth of this laser is less than 1 MHz, with a maximum output power of 30 mW. The driving signal was generated by a waveform generator (WFG, TFG2924A). The laser was passed through attenuators (Att, neutral density filters) to vary the light intensity and then collimated to an AOM, which was also derived by the same waveform generation. The driving signal was a square wave with a repetition frequency of f and a duty cycle of Δ. In this experiment, two different frequencies (*i.e.*, 1 MHz and 100 kHz) and three duty cycles (10%, 20% and 50%) were used, as shown in Figs. S2b and S2c, respectively. The diffraction laser was then expanded by a pair of lenses ($L_1$ and $L_2$) and collimated to the microlens array. At last, the photon statistics of the thermal light were detected by an array of SPCMs (20 SPCMs) and converted to a personal computer through an MCTE with TCSPC function.

## 3. Determination of the photon number probability distributions and high-order correlation functions

In this experiment, each photon was detected by an SPCM, and the arrival time of each photon was determined by the MCET. The pulse widths of the picosecond laser and the super-bunching light source were close to 100 ps. Considering the instrumental response function of the experimental system, the pulse width of the detection signal was about 2 ns. To eliminate the different arriving times of each SPCM originating from the transition time between each SPCM and the difference between channels, we first aligned the arriving times of all SPCMs by additional time delay for each SPCM. Fig. S3a illustrates the schematic of the detected photon sequences for 31 channels. For photon number distribution probability measurement, we counted the number of the photons within 6 ns (*i.e.*, $\Delta t$= 6 ns to ensure that all the photons emitted from the light source were considered) after the trigger from the light source. Then we classified and counted the number of the $N$-photon events, for example, 16-photon events ($PE_{16}$) under the pumping power of 1.85 W and $\langle n \rangle$ of $1.89\times10^{-1}$ with an acquisition time of 11.73 s were 69939. At last, the zero-photon events can be determined by:

$$PE_0 = f \times t - \sum_{N=1}^{31} PE_N \tag{2}$$

where $f$ is the repetition frequency of the light source, and $t$ is the acquisition time. Then, the photon number probability distribution can be demonstrated by:

$$P(N) = \frac{PE_N}{f \times t}, \quad N = 0, 1, \cdots 31 \tag{3}$$

For the measurement of the second-order correlation function, $g^{(2)}(\tau)$, we first counted the 2-photon events with the time delay between photon sequences of two channels to be 0 ns (*i.e.,* $PE_2(t_2\text{-}t_1=\Delta t=0)$), as shown in Fig. S3b. Then we counted the 2-photon events with the time delay between photon sequences of two channels to be $\Delta t_j$ (*i.e.*, $PE_2(t_2\text{-}t_1=\Delta t_j)$), by shifting the time sequences of the channel 2 (Ch-2) with a time of $\Delta t_j$ (Fig. S3c). In this case, $g^{(2)}(\tau)$ can be demonstrated by calculating 2-photon events as a function of time delay. Generally, $g^{(2)}(0)$ can be determined by:

$$g^{(2)}(0) = \frac{PE_2(\Delta t = 0)}{PE_2(\Delta t = mT)} \tag{4}$$

where $PE_2(\Delta t=0)$ and $PE_2(\Delta t=mT)$ are the calculated 2-photon events with the time delay between the two channels, $m$ is an integer ($m\neq0$), and $T$ is the period of the light source ($T=1/f$). For the measurement with tiny $\langle n \rangle$, there will be a significant fluctuation between $PE_2(\Delta t=mT)$, as shown in the inset of Fig. 1c, the mean value of 20 $PE_2(\Delta t=mT)$ will be used for equation (Eq.) 4. For some extreme conditions, the calculated $PE_2(\Delta t=mT)$ and $PE_2(\Delta t\neq mT)$ share similar values, as shown in Fig. S6. In this case, most of $PE_2(\Delta t\neq mT)$ was zero. To determine $g^{(2)}(0)$ in this case, we used the mean value of $PE_2(\Delta t\neq 0)$ ranging from $m$=-10 to $m$=10, expecting the time delay between -50 ns to 50 ns (due to the dead time of the SPCM and thus almost zero counts in this region).

For the measurement of the high-order correlation function, $g^{(N)}(0)$ ($N$>2), we only calculated the $N$-photon events from selected $N$ channels with the time delay of zero and integral periods. Taking $g^{(5)}(0)$ as an example, as illustrated in Fig. S3d, we first counted 5-photon events with the time delay between all channels to be zero (*i.e.*, $PE_5(t_2-t_1=0, t_3-t_1=0, t_4-t_1=0, t_5-t_1=0)$). Then, we shifted the time sequences of the other four channels (Ch-2, Ch-3, Ch-4, and Ch-5) with an integral period, *i.e.*, $m_2T$, $m_3T$, $m_4T$, and $m_5T$, respectively. Here, $m_2$ to $m_5$ are integers while $m_2\neq m_3\neq m_4\neq m_5\neq 0$. Then, we counted 5-photon events with this time delays ((*i.e.*, $PE_5(t_2-t_1=m_2T, t_3-t_1=m_3T, t_4-t_1=m_4T, t_5-t_1=m_5T)$), as shown in Fig. S3e. In this case, $g^{(5)}(0)$ can be determined by

$$g^{(5)}(0)=\frac{PE_5(t_2-t_1=0,t_3-t_1=0,t_4-t_1=0,t_5-t_1=0)}{PE_5(t_2-t_1=m_2T,t_3-t_1=m_3T,t_4-t_1=m_4T,t_5-t_1=m_5T)\big|_{m_2\neq m_3\neq m_4\neq m_5\neq 0}} \quad (5)$$

Figs. S4a and S4b illustrate the result of $g^{(3)}(t_2-t_1, t_3-t_1)$; all the values were normalized by the mean valve of $PE_3(m_2T, m_3T, m_2\neq m_3)$, as highlighted by the green squares. Therefore, $g^{(3)}(0)$ was determined to be about 9314. The values of $g^{(3)}(t_2-t_1=0, t_3-t_1\neq 0)$, or $g^{(3)}(t_2-t_1\neq 0, t_3-t_1=0)$, or $g^{(3)}(t_2-t_1\neq 0, t_3-t_1\neq 0, t_2= t_3)$ can be regarded as $g^{(2)}(0)$, where the mean value was about 72.27.

## 4. Simulation of the Poisson and thermal photon number probability distributions.

The coherent state optical field satisfies the Poisson distribution and can be expressed as:

$$P_c(N)=\frac{\langle n\rangle^N e^{-\langle n\rangle}}{N!}. \quad (6)$$

where $N$ and $\langle n\rangle$ are the number of bundled photons and the mean number of photons per pulse. The thermal field satisfies the Bose-Einstein distribution, which can be expressed as:

$$P_{\mathrm{t}}(N)=\frac{\langle n\rangle^{N}}{(1+\langle n\rangle)^{N+1}}. \tag{7}$$

## 5. Experimental setup of the high-order correlated imaging.

The pump power of the main amplifier for the super-bunching laser was set to 1.85 W, the bandwidth and mean number of the super-bunching laser were optimized by an attenuation (Att) and a band-pass filter (BPF). The light beam was reflected by a glove mirror and focused on the imaging object by a lens ($L_1$). Here, we used three letters, “SXU”, as the target object. The scattering light from the object was collimated by another lens ($L_2$) and expanded by a telescopic system ($L_3$ and $L_4$). Then, the scattering light was collected by a fiber bundle, which was also used for photon number probability distributions, and guided the photons to each SPCM. To ensure enough photons in each SPCM within the integral time (0.1s), we only performed high-order correlated imaging with 20 SPCMs. A thermal light source driven by a noise signal is used so as to simulate real-world conditions with intense stray noise and broad-band illumination. The thermal light overlays the super-bunching laser through a beam splitter (BS) and is detected by the identical SPAD arrays. The $N$th-order correlated imaging is demonstrated by calculating the Nth coincidence events from arbitrary $N$ SPADs, the imaging is determined by scanning the object (the integral times were 0.5s or 0.1s per pixel here).

**Data availability**

All the data supporting the findings of this study are available within this Article and its Supplementary Information. Any additional information can be obtained from the corresponding authors upon reasonable request. Source data are provided in this paper.

**Acknowledgments**

The authors gratefully acknowledge support from the National Key Research and Development Program of China (Grant No. 2022YFA1404201), Natural Science Foundation of China (Nos. U22A2091, 62222509, U23A20380, 62127817, and 62205187), Shanxi Province Science and Technology Innovation Talent Team (No. 202204051001014), the key research and development project of Shanxi Province (202102030201007), and 111 projects (Grant No. D18001).

**Author contributions**

C. C. Q., J. M. L., and L. T. X. conceived and designed the experiment. C. B. Q, Y. Y. L., Y. Y., and X. D. L. conducted the experiments and analyzed the data. Y. R. S., X. D. Z., S. P. H., G. F. Z., R. Y. C., and J. Y. H. contributed to the optical setup, SPCMs array design, and pseudothermal light preparation. Z. C. Y., X. H. L., and S. T. J. contributed to the electronic measurements and data analysis. J. M. L., Z. H. L., and Y. Q. G. contributed to the theoretical analysis and simulations. C. B. Q., J. M. L., and L. T. X. wrote the manuscript with input from all other authors. C. B. Q. and L. T. X. supervised the project.

**Competing interests**

The authors declare no competing interests.

**Additional information**

Supplementary information.